\documentstyle[12pt]{article}
\newcommand{\AmS}{{\protect\the\textfont2
  A\kern-.1667em\lower.5ex\hbox{M}\kern-.125emS}}

\begin{document}

{\LARGE \bf Migration of matter from the Edgeworth--Kuiper and main asteroid  
belts to the Earth }

\centerline{}

\centerline{\bf \Large S.I. Ipatov}

\centerline{\it Institute of Applied Mathematics (Moscow, Russia), NASA/GSFC (USA),}

\centerline{\it ipatov@spp.keldysh.ru, siipatov@hotmail.com}

\centerline{}

Proceedings of IAU Colloquium No 181 and COSPAR Colloquium No. 11 "Dust in the solar system and other planetary systems" (April 10-14, 2000, Canterbury, UK), in press.

\centerline{}


\begin{abstract}
A considerable portion  of
 near-Earth objects could have come from the trans-Neptunian belt. Some of them 
have aphelia deep inside Jupiter's orbit during more than 1 Myr.

\end{abstract}

\section{INTRODUCTION}

The main asteroid belt (MAB), the Edgeworth--Kuiper belt (EKB), and comets belong
to the main sources of dust in the Solar System. 
Most of Jupiter-family comets came from the EKB. 
Comets can be destructed due to close encounters
with planets and the Sun, collisions with small bodies, and internal forces. 
    We support \cite{i99,i01} the Eneev's idea \cite{e80} that the largest objects in the EKB and MAB could be formed
directly by the compression of rarefied dust condensations of the protoplanetary cloud
but not by the accretion of small (for example, 1-km) planetesimals.
The total mass of planetesimals that entered the EKB from the feeding zone
of the giant planets during their accumulation could exceed tens
of Earth's masses $m_\oplus$ \cite{i87,i93}. These planetesimals increased 
eccentricities of 'local' trans-Neptunian objects (TNOs) and swept most of these TNOs.
A small portion of such planetesimals could left beyond Neptune's orbit in highly
eccentric orbits. The
results of previous investigations of migration and collisional evolution of minor bodies 
were summarized in \cite{i00,i01}. Below we present mainly our recent results.

\section{MIGRATION OF MATTER TO A NEAR-EARTH SPACE}

   Asteroids leave the MAB via some regions corresponding to
resonances with Jupiter, Saturn, and Mars. They get into these
regions mainly due to collisions. Gravitational influence of the
largest asteroids plays a smaller role. The number of resonances delivering
bodies to the Earth is not small (more than 15) \cite{mn99}. 
So even due to small variations in semimajor axes $a$, some asteroids can get into the 
resonances, and the role of mutual gravitational influence of 
asteroids in their motion to the Earth may not be very small. Small bodies can get into
the resonant regions also due to the Yarkovsky orbital drift. For dust
particles we also need to take into account the Pointing--Robertson effect, 
radiation pressure, and solar wind drag.

Objects leave the EKB mainly due to the gravitational influence of
planets \cite{d95}. During last 4 Gyr several percents of TNOs could change 
$a$ by more than 1 AU due to the gravitational interactions with
other TNOs \cite{i01}. For most of other TNOs such variations in $a$ were 
less than 0.1 AU. The role of mutual gravitational
influence of TNOs in evolution of their orbits may be greater than
that of their collisions. Even small variations in orbital elements of TNOs
due to their mutual gravitational influence and collisions can
cause large variations in orbital elements due to the
gravitational influence of planets. TNOs can leave the EKB 
(and comets leave the Oort cloud) without
collisions. Therefore, some cometary objects migrating inside the Solar System can be
large. The largest objects (with $d$$\ge$$10$ km) that collided the Earth 
during last 4 Gyr could be mainly of cometary origin.

We investigated the evolution for intervals $T_S$$\ge$5
Myr of 2500 Jupiter-crossing objects (JCOs) under the 
gravitational influence of all planets, except for Mercury 
and Pluto (without dissipative factors). In the first series 
we considered $N$=2000 orbits near the orbits of 30 real 
Jupiter-family comets with period $<$$10$ yr, and in the 
second series we took 500 orbits close to the orbit of Comet 
10P Tempel 2 ($a$$\approx$3.1 AU, $e$$\approx$0.53, $i$$\approx$$12^\circ$). 
We calculated the probabilities of collisions 
of objects with the terrestrial planets, using orbital 
elements obtained with a step equal to 500 yr and then 
summarized the results for all time intervals and all 
bodies, obtaining the total probability $P_\Sigma$ of 
collisions with a planet and the total time interval 
$T_\Sigma$ during which perihelion distance $q$ of bodies was 
less than a semimajor axis of the planet. The values of $P_r $$=$$10^6 P$$=$$ 
10^6 P_\Sigma /N$ and $T$$=$$T_\Sigma /N$ are presented 
in the Table together with the ratio $r$ of the total time 
interval when orbits were of Apollo type (at $a$$>$1 AU, $q$$=$$a(1-e)$$<$1.017 AU, $e$$<$$0.999$) 
to that of Amor type ($1.017$$<$$q$$<$1.33 AU); $r_2$ is the same as 
$r$ but for Apollo objects with eccentricity $e$$<$0.9. 
For observed near-Earth objects (NEOs) $r$ is close to 1. 

\vspace{3mm}

 {\bf Table:} Values of $T$ (in kyr), $T_c$$=$$T/P$ (in Myr), $P_r$, $r$, $r_2$ for the terrestrial planets

$ \begin{array}{llccccccccc} 

\hline

  &  & $Venus$ & $Venus$ & $Earth$ & $Earth$ & $Earth$ & $Mars$ & $Mars$ & - & - \\

\cline{3-9}

 & N & T & P_r & T & P_r & T_c & T & P_r & r & r_2 \\

\hline

$JCOs$ & 2000 & 9.3 & 6.62 & 14.0 & 6.65 &2110 & 24.7 & 2.03 & 1.32 &1.15\\
$comet $ 10P & 500 & 24.9 & 16.3 & 44.0 & 24.5 & 1800 & 96.2 & 5.92 & 1.49 &1.34\\
3:1 $ reson.$& 144  &  739  & 529 & 1227  &  626 & 510 &  2139 & 116& 2.05 & 1.78\\
5:2 $ reson.$& 144  & 109 & 54.5  & 223  & 92.0 & 416 & 516 & 19.4& 1.28& 1.15 \\

\hline
\end{array} $ 

\vspace{3mm}

For integrations we used the Bulirsh-Stoer method (BULSTO) and a symplectic method. 
The probabilities of collisions of former JCOs with planets were close for these 
methods, but bodies got resonant orbits more often in the case of BULSTO. Besides 
JCOs, we considered asteroids with initial values of $e$ and $i$ equal to 0.15 and 
$10^{\circ}$, respectively. For the asteroids initially located at the 3:1 resonance 
with Jupiter, we found that the ratio $r_{hc}$ of the number of asteroids ejected 
into hyperbolic orbits to that collided with the Sun was 5.6 for BULSTO and 0.38 
and 0.87 for a symplectic method for a step of integration equal to 10 and 30 days, 
respectively. So in some cases a symplectic method can give a large error. For the 
5:2 resonance with Jupiter, $r_{hc}$ equaled 20 and 10 for BULSTO and symplectic 
methods, respectively. In the Table for asteroids  we present  only results 
obtained by the BULSTO code at $T_S$$=$50 Myr (at $T_S$$=$10 Myr the values of $P$ and 
$T$ are smaller by a factor less than 1.2 and 1.01 for the 3:1 and 5:2 resonances, 
respectively) and for TNOs we present results obtained by both codes.

The total time during which former 2000 JCOs 
were in Apollo-type and Amor-type orbits was 28.7 and 21.75 
Myr, respectively, but 12.7 and 11.4 Myr of the above times 
were due to three objects. 
We found several former TNOs that moved for more than 1 Myr in orbits with 
aphelion distance $Q$$<$4.7 AU. The time interval during which a body had $Q$ 
less than 3.2 and 3.7 AU exceeded 0.1 and 2.6 Myr, respectively. 

Most of the collisions of former JCOs with the Earth were from orbits with aphelia 
inside Jupiter's orbit. The probability of collisions with the Earth for 3 former 
JCOs, each of which moved for more than 1 Myr in Earth-crossing orbits (mainly with 
$Q$$<$4.7 AU) was 1.5 times greater than that for the other 1997 JCOs. About 1 of 
300 JCOs collided with the Sun. In \cite{ih99} we considered a much smaller number of 
objects, which didn't get aphelia inside Jupiter's orbit and the values of $P_r$ and 
$T$ were smaller than those in the Table.
For 2000 JCOs we consider, the mean probability of collisions with Venus is about 
the same as with Earth, and that with Mars is smaller by a factor of 3. These 
values are mainly due to a few bodies that moved during more than 1 Myr in orbits 
with aphelia deep inside Jupiter's orbit (for such bodies usually more than 80\% 
of collisions with planets were from orbits with $Q$$<$4.2 AU). If we consider 
1000 JCOs, for which most of the collisions with planets were from orbits with 
$Q$$>$4.2 AU, then the mean probability for Venus and Mars is less by a factor 
of 1.6 and 3, respectively, than that for Earth. Therefore, the ratio of the total 
mass of icy planetesimals that migrated from the feeding zone of the giant planets 
and collided with the planet to the mass of this planet was greater for Mars than 
that for Earth and Venus. 

     The mean time during which an object crossed Jupiter's orbit was 0.13 
Myr for 2500 JCOs. An object had period $P_a$$<$10 yr usually only during 
about 12\% of this time, so we think that our consideration of initial objects 
with only $P_a$$<$10 yr does not influence much on the obtained results. 
At $N$$=$2000 for $10$$<$$P_a$$<$20, $20$$<$$P_a$$<$50, $50$$<$$P_a$$<$200 
yr, we got 23\%, 22\% and 16\%, respectively. One former JCO spent some time in 
orbits with aphelia deep inside Jupiter's orbit, and then it moved for tens of 
Myr in the trans-Neptunian region, partly in low eccentricity and partly in high 
eccentricity orbits. This result shows that some bodies can get from the MAB 
into the trans-Neptunian region, and that typical TNOs can become scattered 
objects (with high eccentricities) and vice versa. 

\section{COLLISIONS WITH THE EARTH}

     The number of TNOs migrating to the inner regions of the Solar System can 
be evaluated on the basis of simple formulas and the results of numerical 
integration. Let $N_J$$=$$p_{JN} P_N N_{TNO}$ be the number of former TNOs with 
$d$$>$$D$ reaching Jupiter's orbit for the given time span $T_{SS}$, where 
$N_{TNO}$ is the number of TNOs with $d$$>$$D$; $P_N$ is the fraction of TNOs 
leaving the EKB and migrating to Neptune's orbit during $T_{SS}$; and $p_{JN}$ 
is the fraction of Neptune-crossing objects which reach Jupiter's orbit for 
their lifetimes. Then the current number of Jupiter-crossers that originated 
in the zone with $30$$<$$a$$<$50 AU equals $N_{Jn}$$=$$N_J \Delta t_J/T_{SS}$, 
where $\Delta t_J$ is the average time during which the object crosses Jupiter's 
orbit. According to \cite{d95}, the fraction $P_N$ of TNOs that left this zone 
during $T_{SS}$$=$4 Gyr under the influence of the giant planets is 0.1-0.2 and 
$p_{JN}$$=$0.34. As mutual gravitational influence of TNOs also takes place \cite{i01}, 
we take $P_N$$=$0.2. Hence, at $\Delta t_J$$=$0.13 Myr and $N_{TNO}$$=$$10^{10}$ 
($d$$>$1 km), we have $N_{Jn}$$=$$2\cdot10^4$. The number of former TNOs now moving 
in Earth-crossing orbits equals $N_E$$=$$N_{Jn}T/ \Delta t_J$. The characteristic 
time $T_{cN}$ between collisions of former TNOs with the Earth is $T /(N_{Jn} P)$. 
For $T$$=$0.014 Myr and $\Delta t_J$$=$0.13 Myr, we have $N_E$$=$2150 and 
$T_{cN}$$\sim$0.1 Myr. $N_E$ is larger than the estimated number $N_{Ee}$ of Earth-crossers 
with $d$$>$1 km (750), and $T_c$$=$$T/P$ is larger than the characteristic time 
$T_{co}$$\approx$100 Myr elapsed till a collision with the Earth
obtained for fixed orbits of the observed NEOs. Such difference 
can be caused by the fact that it is difficult to observe NEOs with high $e$ and 
$i$ and $N_{Ee}$ doesn't include such NEOs. It may be also probable that the number 
of 1-km TNOs is smaller than $10^{10}$. As comets can get NEO and asteroidal orbits,
a considerable portion of dust produced  by NEOs and even some dust produced in 
the MAB can be of comet origin.

     The total mass of water delivered to the Earth during the formation of the 
giant planets is $M_w$$=$$M_J P_{JE} k_i$, where $M_J$ is the total mass of 
planetesimals from the feeding zones of these planets that became Jupiter-crossers
during evolution, $P_{JE}$ is a probability $P$ of a collision of a former JCO with 
the Earth during its lifetime, and $k_i$ is the portion of water ices in the 
planetesimals. For $M_J$$=$$100m_\oplus$, $k_i$$=$0.5, and 
$P_{JE}$$=$$6.65\cdot 10^{-6}$, we have $M_w$$=$$3.3\cdot10^{-4}m_\oplus$. 
This value is greater by a factor of 1.5 than the mass of the Earth oceans. 
The mass of water delivered to Venus can be of the same order of magnitude 
and that delivered to Mars can be less by a factor of 3. Some TNOs with 
$a$$>$50 AU can also migrate to the orbits of Jupiter and Earth. Collisions 
of comets with small bodies and nongravitational forces can decrease $Q$. 
Asher {\it et al.} \cite{a00} showed that the rate at which objects may be 
decoupled from Jupiter and attain orbits like NEOs is increased by a factor 
of four or five, if nongravitational forces are included as impulsive effects. 
So the values of $P_r$ and $T$ can be larger than those in the Table.
Rickman {\it et al.} \cite{r01} also concluded that comets play an important 
role among all km-sized impactors.
As it is easier to destroy icy bodies than stone or metal bodies, the portion 
of TNOs among NEOs for bodies 
with $d$$<$100 m may be greater than that for 1-km bodies.
In future, when people will make settlements on the Moon and terrestrial planets, 
small icy comets can be move by rockets to the orbits around these celestial 
bodies in order to be sources of water.

This work was supported by INTAS (00-240), RFBR (01-02-17540), and NASA (NAG5-10776).


\begin{thebibliography}{99}
\bibitem{a00} D.J. Asher, M.E. Bailey, and D.I. Steel, In "Collisional Processes in 
the Solar System",  ed. by M.Ya. Marov and H. Rickman, ASSL V. 261 (2001) 121-130.
\bibitem{d95} M.J. Duncan, H. F. Levison, and S. M. Budd, Astron. J. 110 (1995) 3073-3081. 
\bibitem{e80} T.M. Eneev, Sov. Astron. Letters 6 (1980) 295-300 in Russian edition.
\bibitem{i87} S.I. Ipatov, Earth, Moon, and Planets 39 (1987) 101-128.
\bibitem{i93} S.I. Ipatov, Solar System Research 27 (1993) 65-79.
\bibitem{i95} S.I. Ipatov, Solar System Research 29 (1995) 261-286.
\bibitem{i99} S.I. Ipatov, Celest. Mech. Dyn. Astron. 73 (1999) 107-116.
\bibitem{i00} S.I. Ipatov, "Migration of celestial bodies in the Solar System", 
Editorial URSS, Moscow, (2000) (in Russian), 320 P.
\bibitem{i01} S.I. Ipatov, Advances in Space Research 28 (2001) 1107-1116. 
\bibitem{ih99} S.I. Ipatov and G.J. Hahn, Solar System Research 33 (1999) 487-500.
\bibitem{mn99} A. Morbidelli and D. Nesvorny, Icarus 139 (1999) 295-308.
\bibitem{r01} H. Rickman, J.A. Fernandez, G. Tancredi, J. Licandro, In "Collisional Processes in the Solar System", 
ed. by M.Ya. Marov and H. Rickman, ASSL V. 261 (2001) 131-142.

\end{thebibliography}
\end{document}